\documentclass[runningheads]{llncs}
\usepackage[T1]{fontenc}
\usepackage{cite}
\usepackage{graphicx}
\begin{document}
\title{Optimized User Experience for Labeling Systems for Predictive Maintenance Applications}
\titlerunning{Optimized User Experience for Labeling Systems}
%
\author{Michelle Hallmann\inst{1}\orcidID{0009-0002-9600-7575} 
\texttt{michelle.hallmann@hshl.de}
\and
Michael Stern\inst{1}\orcidID{0009-0002-7558-5384} \and
Francesco Vona\inst{1}\orcidID{0000-0003-4558-4989}\and
Ute Franke\inst{2}\orcidID{0009-0009-3945-4775} \and
Thomas Ostertag\inst{3}\orcidID{0009-0005-8617-3256 } \and
Benjamin Schlüter\inst{4}\orcidID{0009-0005-0374-363X} \and
Jan-Niklas Voigt-Antons\inst{1}\orcidID{0000-0002-2786-9262}
}
 
\authorrunning{Hallmann et al.}
 
\institute{Hamm-Lippstadt University of Applied Sciences,  59557 Lippstadt, Germany \and
5micron GmbH, Rudower Ch 29, 12489 Berlin, Germany \and
OSTAKON GmbH, Forster Str. 54, 10999 Berlin \and
Deutsche Eisenbahn Service AG, Pritzwalker Str. 8, 16949 Putlitz}
 
\maketitle              

\noindent
This version of the contribution has been accepted for publication, after peer review. The Version of Record is available online at: https://doi.org/10.1007/978-3-031-76821-7\_4
 
\begin{abstract}
This paper presents the design and implementation of a graphical labeling user interface for a monitoring and predictive maintenance system for trains and rail infrastructure in a rural area of Germany. Aiming to enhance rail transportation's economic viability and operational efficiency, our project utilizes cost-effective wireless monitoring systems that combine affordable sensors and machine learning algorithms. Given that a successful labeling phase is indispensable for training a supervised machine learning system, we emphasize the importance of a user-friendly labeling user interface, which can be optimally integrated into the daily work routines of annotators. The labeling system has been designed based on best practices in usability heuristics and will be validated for usability and user experience through a study, the protocol for which is presented here. The value of this work lies in its potential to reduce maintenance costs and improve service reliability in rail transportation, contributing to the academic literature and offering practical insights for research on effective labeling user interfaces, as well as for the development of labeling systems in the industry. Upon completion of the study, we will share the results, refine the system as necessary, and explore its scalability in other areas of infrastructure maintenance. 

\end{abstract}

\section{Introduction}
The economic success of rail transportation is heavily dependent on infrastructure and vehicle maintenance costs \cite{ref_Maintenance}. Insufficient maintenance of rail infrastructure and vehicles provokes train delays and malfunctions \cite{ref_TrainDelays}, leading to heightened public discontentment with the public transportation system \cite{ref_ValuationReliabiltyTravel,ref_UnreliablityPublicTransport}. Therefore, cost-efficient solutions must be found to enhance transport capacity in rural areas and increase the attractiveness of rail transportation.
A promising solution involves deploying cost-effective wireless monitoring systems for rail infrastructure and vehicles \cite{ref_DegradationPrediction}, combining affordable sensors and machine learning algorithms for data analysis and predictive maintenance. These systems must be designed with low-threshold retrofit solutions to allow seamless integration into existing infrastructure.  
To build a reliable predictive maintenance system based on machine learning algorithms, a high-quality labeling process must be ensured, as the capabilities of a machine learning system are heavily dependent on the quality of the training data \cite{ref_DataQualityML}. The labeling, which is needed for a machine learning system to be deployed, is often seen as a tedious task \cite{ref_TediousLabeling}, which can when the annotator is using a system with low usability and user experience can lead to frustration \cite{ref_UXPerformance}, and thus lower performance results and lower label quality. Thus, ensuring that the user of the labeling system inputs high-quality data necessitates a user interface with a high usability and user experience, making the labeling activity pleasant instead of exhausting or annoying. 
Despite the importance of high usability and user experience in labeling systems, few guidelines exist for designing labeling user interfaces for predictive maintenance, with even fewer specific examples presented for software solutions \cite{ref_LabelUXGuidelines}. 
This paper presents the design of a graphical labeling user interface for integrating a monitoring and predictive maintenance system for trains and rail infrastructure in a rural area of Germany.  It is part of a research project where we assess the effectiveness of structure-borne noise measurement methods for monitoring and maintenance investigations on rails and vehicles. For this purpose, two train cars and one rail section are equipped with structure-borne noise measurement sensor systems for data collection from 5micron, which are general-purpose, cost-effective, and highly scalable.
Acoustics' special characteristic of deriving specific findings from non-specific signals is used to detect the condition and material properties of vehicles and infrastructure. The two sensors in the train cars monitor the state of the rails, while the sensor on the rail section monitors the state of the train cars as they enter and leave the railway workshop. 
The train drivers label events occurring during the train rides, while the workshop foreman labels faults and repairs to the vehicles. A distributed ledger network (DLT) is implemented, which lifts the infrastructure to the web3 standard and makes the essential use cases cross-company-wide available. The DLT network realizes the decentralized data exchange and transfers the labeling data to a data analysis server, where machine learning algorithms process the labels and training data to generate recommended measures for the workshop foreman. With the presentation of the processes and the design implementation in our project, we aim to provide suggestions for the design of labeling systems in the area of Industry 4.0, specifically focusing on the integration of predictive maintenance systems in the rail transportation industry. Our design decisions are based on the best practices in usability heuristics \cite{ref_UsabilityEngineering}. Further, the study protocol we plan to conduct as a next step for evaluating our labeling user interface's usability and user experience is also presented.

\section{Related Work}
\subsection{Supervised Machine Learning for Predictive Maintenance}
If machines are not serviced promptly in the event of imminent defects, this can result in major damage and, therefore, higher repair costs, which are detrimental to the company. However, if systems are serviced too regularly and without need, time and economic resources are used that are not yet required \cite{ref_ImplementationPM}. With machine learning for predictive maintenance, developers can build monitoring systems out of sensors, which detect anomalies in usage and predict future damage and repair measures to improve the maintenance of machinery \cite{ref_MachineLearning}.
To develop predictive maintenance systems in this field, usually supervised learning is required to train the machine learning system, where in addition to the sensor data, experts from the field assign labels to the data \cite{ref_MLFuture}. The success of the machine learning system depends on the accuracy and quality of the data and labels, which is why the acquisition of data and labels in the training phase is an important part of the process \cite{ref_MLPM}. 

\subsection{Usability and User Experience for Labeling Quality}
In the conception, design, and development of user interfaces, two concepts are of great importance and must be considered by designers and developers throughout: Usability and User Experience. Usability focuses on making the system simple, intuitive, and user-friendly, ensuring users can achieve their goals with the system \cite{ref_UsabilityUX}. As early as 1994, Nielson \cite{ref_UsabilityEngineering} developed 10 heuristics to help systems achieve better usability. These heuristics include, for example, making the system status visible to the user when designing the system, maintaining consistency and adhering to standards, and displaying only necessary elements through aesthetic and minimal design.
User Experience, as described by Jacobsen and Meyer \cite{ref_UsabilityUX}, encompasses the comprehensive overall experience that the user has before, during, and after using the system. Ideally, this experience should leave the user happy or enthusiastic, encouraging them to return to the system  \cite{ref_UsabilityUX}. User experience considers the user's perception and emotional experience during use, as well as their needs and the degree to which these needs are fulfilled  \cite{ref_UsabilityUXDesign,ref_InteraktiveSysteme}.
A system's degree of usability and user experience indicates how much users like and enjoy interacting with this system \cite{ref_UsabilityUX,ref_InteractionDesign}. The labeling task can be very tedious  \cite{ref_TediousLabeling}, but as this part of the process and the label quality are very important, we aim for a high Usability and User Experience of the Labeling User Interface to ensure efficient and high-quality labeling by the expert.

\subsection{Heuristics and Guidelines for Labeling User Interfaces}
To our knowledge, there are few guidelines in the literature on designing user interfaces for labeling systems. The only guideline we found specifically for this kind of system was the one from Passos developed in 2021 \cite{ref_LabelUXGuidelines}, where the guidelines were derived from the usability heuristics from Nielson \cite{ref_UsabilityEngineering}. As the guidelines were not scientifically validated yet, we decided to base our design decisions on the usability heuristics from Nielson, which serve as general guidance for the design of user interfaces \cite{ref_UsabilityEngineering,ref_UsabilityUX,ref_UsabilityUXDesign,ref_InteraktiveSysteme}. Thus, we will explain our design decisions based on Nielson's usability heuristics when we present the Labeling User Interface.

\section{Software Design}
\subsection{Conceptualization of the Labeling Process}
To develop an effective integrable labeling system, we conceptualized the labeling processes in collaboration with the railroad company, sensor system developers, and data exchange companies. To this end, we held several meetings involving employees from the train company. Rail infrastructure experts informed us about the previous maintenance processes and common messages during journeys. During a visit to the workshop, we consulted with the workshop foreman about the previous maintenance processes for the train cars and collaboratively discussed the design of the labeling processes. The labeling processes were designed closely with the train company employees and the entire project consortium.
As recommended in Guideline 1, “Identification of the labeling task,” the labeling requirements were identified during the discussions with the sensor company. Even if the data type “structure-borne sound recording” to be labeled is not listed among the data types in the guideline, we could determine the multi-label text classification as the labeling task since descriptions should be assigned to the fault.

\subsection{Labeling Environment}
We have decided to assign the task of labeling rail faults to the train drivers and faults on train cars to the workshop foreman, who possesses the necessary technical expertise. To ensure that these employees' involvement in the research project does not disrupt their daily work, the user interface (UI) for labeling must be seamlessly integrated into their work processes.
The workshop foreman handles maintenance orders on the computer in his office within the workshop. Therefore, we selected the computer as the medium for labeling in collaboration with him. He confirmed he could keep a website open alongside the maintenance system, allowing him to perform labeling tasks between assignments.
Each driver is equipped with a company cell phone and a company tablet. The drivers’ supervisor assured us that they could dedicate a few minutes at the end of their shifts to label events that occurred during their routes. Since the service tablet’s larger size can accommodate more information than the service cell phone, we selected the service tablet as the medium for labeling infrastructure events. To create a simple system, we used a website as the UI. After logging in, users are directed to either the train drivers' labeling interface web page or the workshop foreman's labeling interface web page.

\subsection{Labeling Sequence}
To label faults on the rails, the train driver is instructed to press a button integrated with the sensor while driving if they observe an anomaly in the rail infrastructure. The sensor data for the corresponding route section is marked and transmitted to the data-analysis server, where it is stored in a database. After completing the train ride, the train driver is prompted to access the labeling user interface on their tablet, where the unlabeled events are listed. 
When a train car enters the workshop, the data from the sensor is forwarded to the server system and embedded as an event on the website. The workshop foreman is instructed to label the found and repaired defects of the inspection by selecting fault categories from predefined lists or creating new ones. 
Once verified and submitted, the selected labels from the train drivers and the workshop foreman are sent to the server, and the system is trained with the newest acquired data.
After recording the labeling system requirements, we created, iterated, and tested initial design approaches. Ultimately, we developed a labeling system that can label faults on the rails and train cars, facilitating the collection of the necessary data for data analysis.

\subsection{Design Methodology}
The design of the labeling system followed a codesign approach, resulting in the formalization of all functional and non-functional requirements for the UI. Following discussions with the entire project team, the design and development team established all functional and non-functional requirements for the user interface intended for the annotators, coordinating these requirements with all stakeholders. Wireframes were used to determine how the elements and functions of the labeling system should be arranged on the screen. Once agreement was reached on the structure of the user interfaces, an initial digital design prototype for the user flows for both the train drivers and the workshop foreman was created.
The labeling system was to be implemented using Vue.js, focusing on developing a minimalist design to ensure the interface was easy to understand and use for the annotators \cite{ref_MinimalismUI}. The Vue.js framework Vuetify \cite{ref_Vuetify} was utilized for this minimalist design. After the first iteration and discussions on the prototype with university experts, additional design iterations were made, leading to the development of the website's final design. 

\section{Labeling User Interface}
The labeling system is divided into a log-in page for all annotators and one dashboard for each labeling use case. For a minimalist design, as recommended in the heuristic "Aesthetic and Minimalist Design" \cite{ref_UsabilityEngineering}, we used white- and gray-colored backgrounds, black font color and the font-family "Roboto". The muted blue tone from the project logo is reused for highlighting and some buttons so that the user recognizes the project when opening the labeling system, aligning with the heuristic "Recognition Rather Than Recall" \cite{ref_UsabilityEngineering}. Our conversations with the railroad employees gave us an impression of their language. As described in the heuristic “Match Between The System And The Real World” \cite{ref_UsabilityEngineering}, we were able to adapt the system to the user and thus simplify its use.

\subsection{Labeling of Train Cars}
\begin{figure}
    \centering
    \includegraphics[width=1\linewidth]{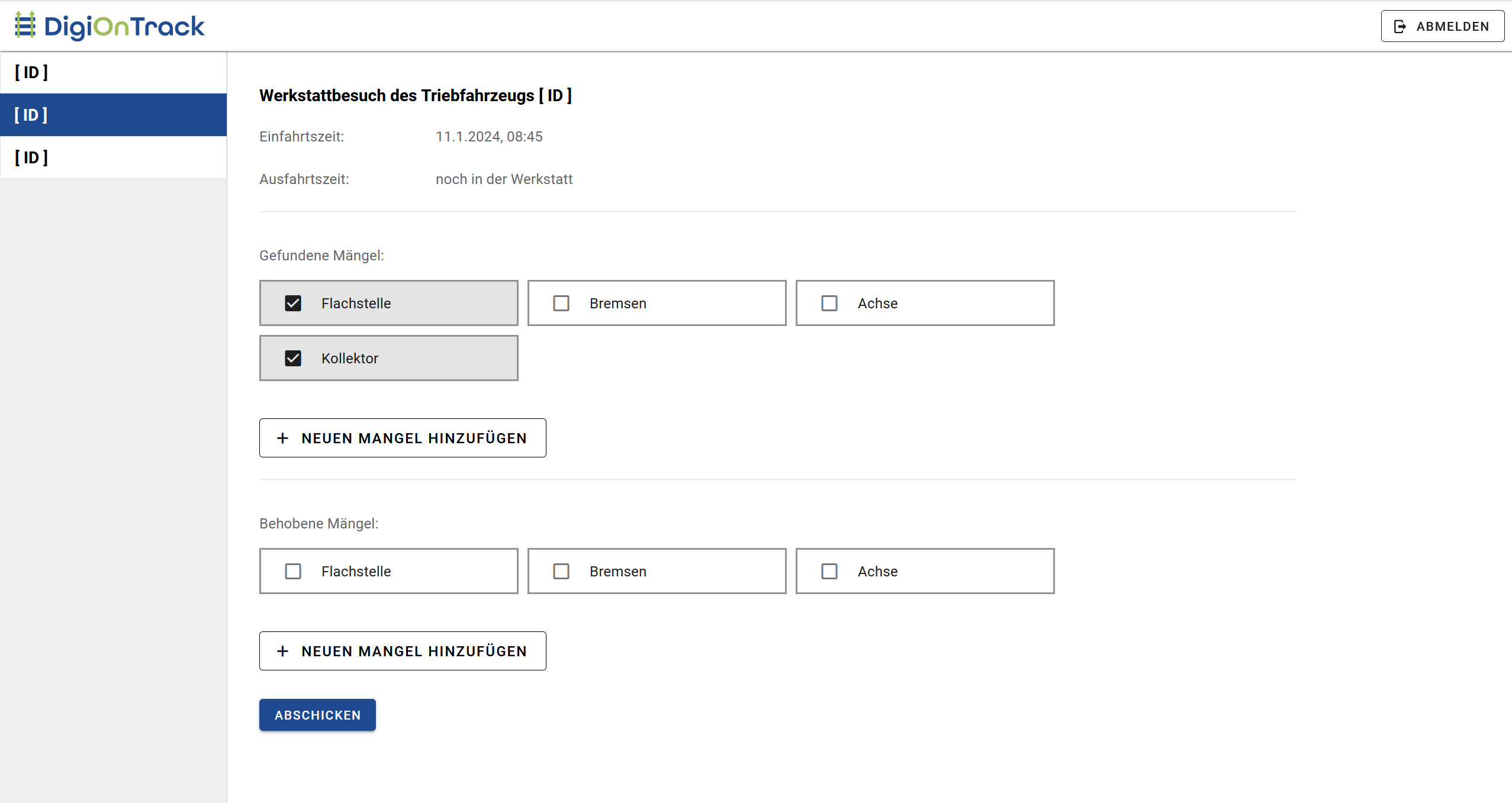}
    \caption{Dashboard for the Labeling of Train Cars for the Workshop Foreman: List with Train Vehicles on the Left, Entry and Exit Time at the Top, Lists with Labels for Labeling of Faults and Repairs in the Center}
    \label{fig:dashboard-trainCarLabeling}
\end{figure}

The user interface for the train car labeling \ref{fig:dashboard-trainCarLabeling} is divided into 3 parts: a bar at the top with the logo and a logout button, a list of events on the left which shows the identification numbers of the train cars which entered and left the workshop and a main view for the labeling. The currently selected event is highlighted in the events list so the user can always see which train car is being labeled.  This aligns with the heuristic “Visibility of System Status” \cite{ref_UsabilityEngineering}. The upper part of the main view provides only the necessary information about the train car, including the train identification number, entrance time, and exit time, keeping the interface minimalist.

\begin{figure}
    \centering
    \includegraphics[width=1\linewidth]{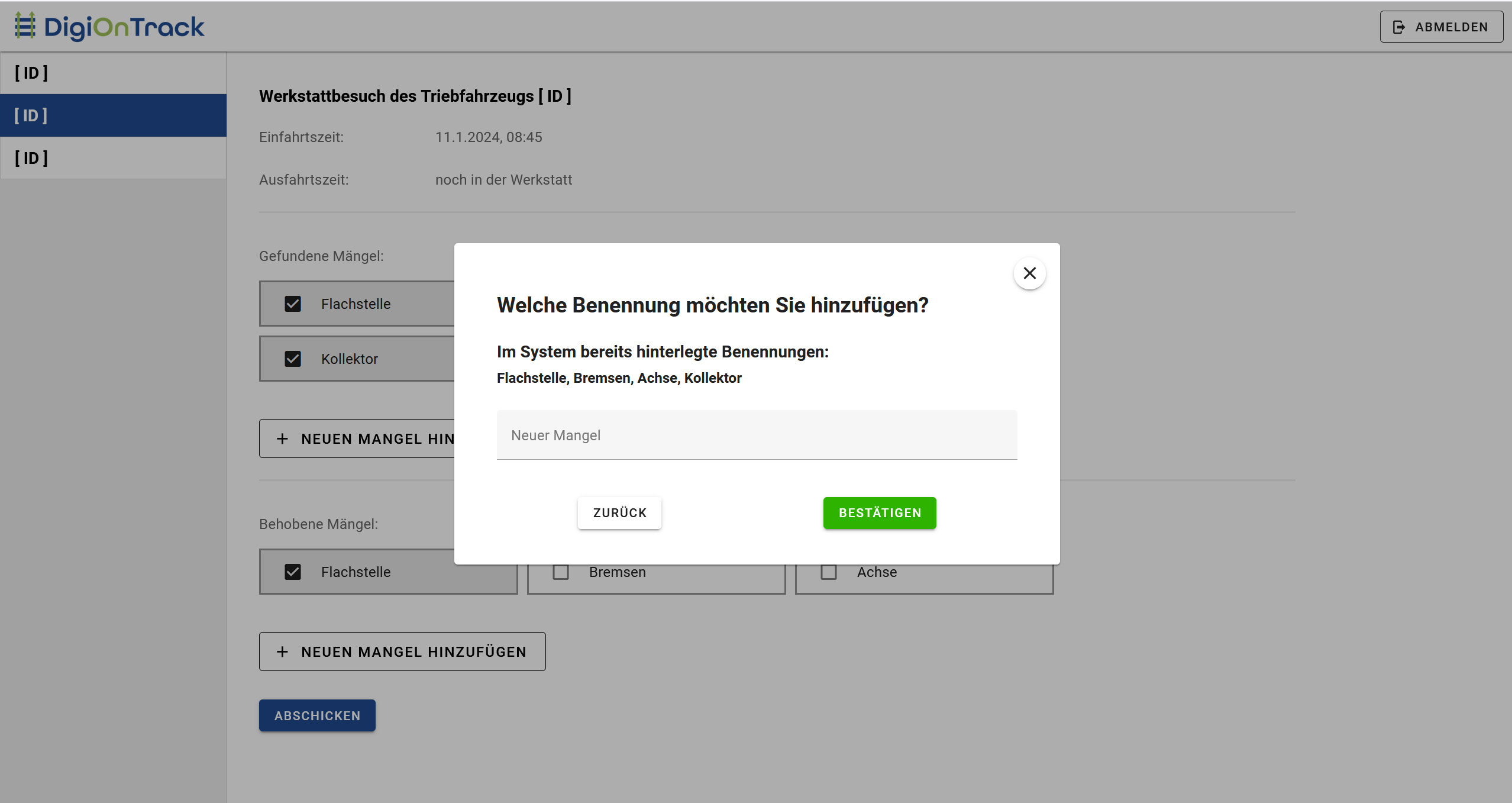}
    \caption{Overlay for Creating a new Label: User is asked for the new Label, List with already available Labels and Input Field for Entering new Label}
    \label{fig:create-label}
\end{figure}

A hierarchical structure for the labels was deemed unnecessary, as there were only three labels at the beginning of the labeling phase, and the workshop foreman expected that not many labels would be added in the future. This decision helps avoid long interaction sequences with the system and keeps the interface minimalist. However, since annotators may need to create new labels, an option to create labels through a button below the list of labels was implemented. As multiple faults can be found or repaired on train cars, the system allows for the selection and creation of multiple labels. 
When the workshop foreman clicks on the button "create new label," a small and simple overlay appears, where he is asked which label he wants to create and can create it through an input field \ref{fig:create-label}. Above the input field, all already selectable labels of the labeling type are listed, so the workshop foreman has an overview of the system while editing it. This might prevent creating a label from the annotator, which already exists in the system and aligns with the heuristic "Error Prevention" \cite{ref_UsabilityEngineering}. The workshop foreman can close the overview through the close button, go back with the back button, or confirm the label to be created by the confirm button. Not only does the close button align with the heuristic "Consistency and Standards," but it also gives the user two easy ways to go back to the labeling if they clicked on the button by mistake, aligning with the heuristic "User Control and Freedom."

\begin{figure}
    \centering
    \includegraphics[width=1\linewidth]{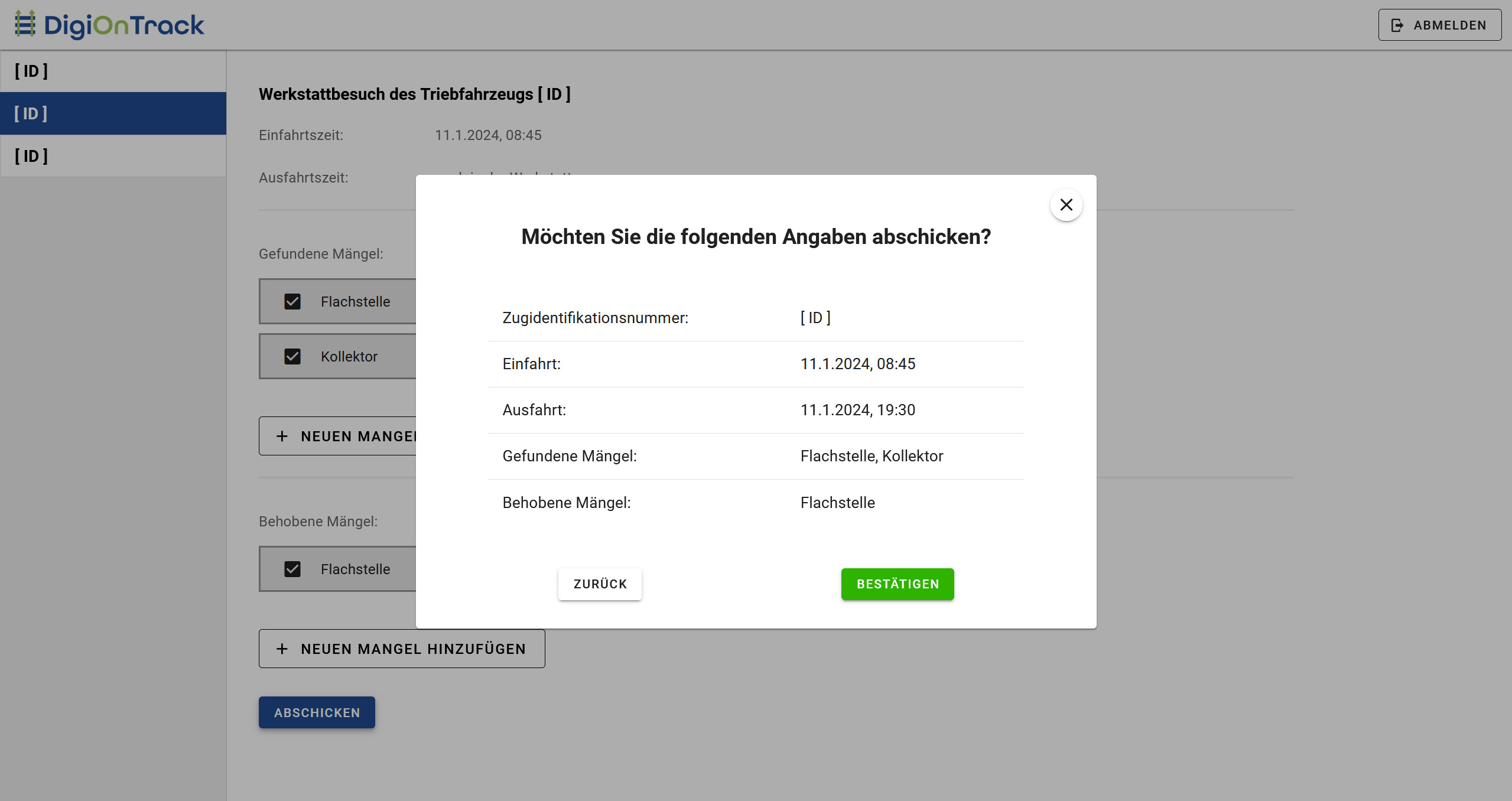}
    \caption{Data Verification before submitting Labels: User is asked if following Data should be submitted, Train Vehicle Data with selected Labels is listed}
    \label{fig:data-verification}
\end{figure}

When the workshop foreman selects the labels and clicks on the blue send button at the bottom of the page, an overlay appears, where the workshop foreman can check the data that he is submitting \ref{fig:data-verification}. This provides another opportunity to correct any mistakes and makes the system status visible by showing which data has been saved and will be sent upon submission \cite{ref_UsabilityEngineering}. The overlay is designed to be simple and apportioned to maintain a minimalist approach. The workshop foreman can close the overlay with a close button, return to the labeling UI with the back button, or submit the data to the server with the confirm button. This again gives the user full control and freedom over what to do next \cite{ref_UsabilityEngineering}.

\subsection{Labeling of Rail Infrastructure}
\begin{figure}
    \centering
    \includegraphics[width=0.5\linewidth]{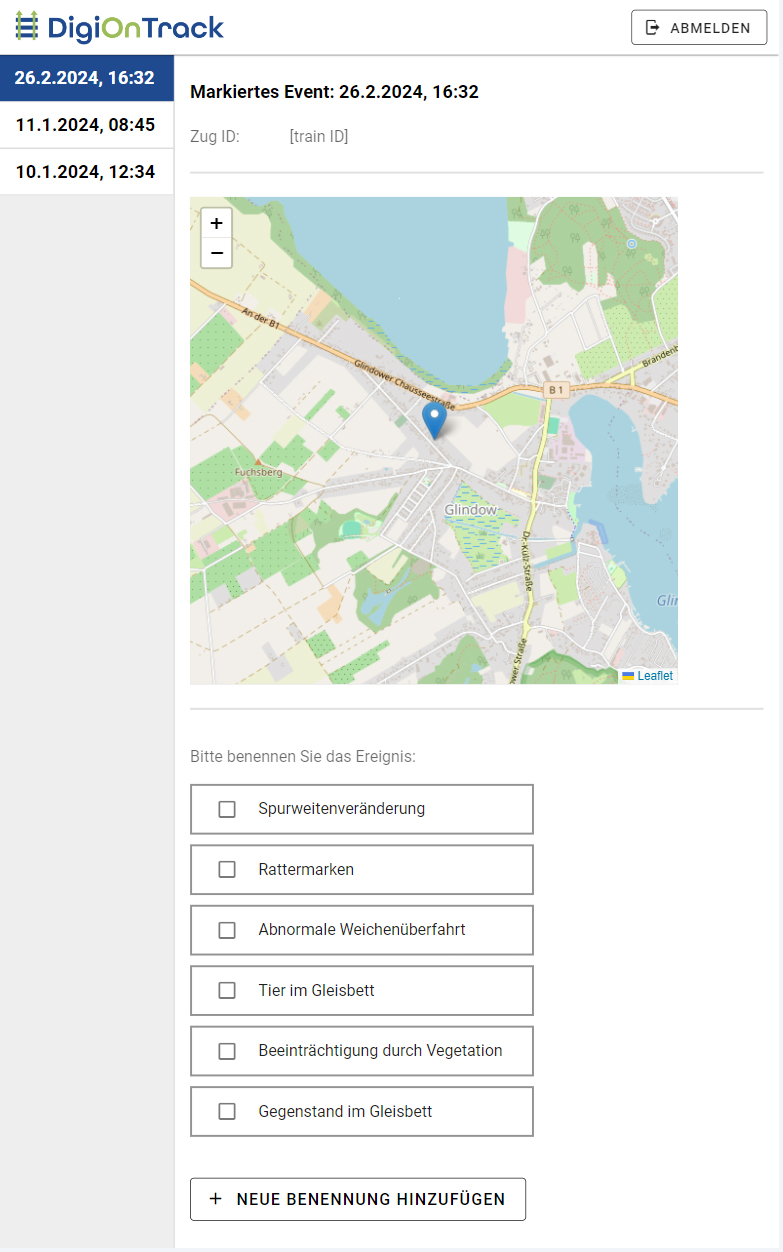}
    \caption{User Interface for Labeling of the Rails for the Train Drivers: On the Left is the List of Events, at the Top Date, Time, Train Identification and Event Location are shown, Below the Map the List with eligible Labels is positioned}
    \label{fig:rails-labeling}
\end{figure}

The user interface for labeling the rails (Figure \ref{fig:rails-labeling}) is similarly constructed to the user interface for train car labeling but is designed for the tablet used by train drivers. On the left side of the screen, events are listed with their event date and time. At the top of the screen is a logo for identification and a logout button to return to the main system, providing recognition, user control, and freedom \cite{ref_UsabilityEngineering}.
In the main part of the screen, the selected event is displayed with its date, time, train identification, and location, ensuring visibility of the system status \cite{ref_UsabilityEngineering}. The location is shown via a pin on a map, implemented using the open-source framework OpenStreetMap \cite{ref_OpenStreetMap}. The pin indicates where the train driver pressed the button to tag an event during the journey. The train driver can zoom in and out and navigate the map by touch to better orient themselves with the location.
Initially, the labels are implemented with a flat hierarchy similar to those in the train car labeling interface. However, if numerous new labels are created, resulting in an overly large list, a hierarchical structure for the labels will be considered for better organization and overview.

\section{Usability and User Experience Study}
In the following, we present the study protocol for a planned usability and user experience study for our developed labeling system to verify these user interface aspects and provide a validated example of labeling the user interface for predictive maintenance for other researchers and developers. Further, we aim to investigate the correlation between the participants' affinity for technological interaction and gender and the application's usability and user experience.

\subsection{Participants}
Due to the challenges of recruiting sufficient specialists for a robust study, we will initially conduct a preliminary study involving academic staff and students from our university. We aim to recruit 40 participants, ensuring diversity in age and gender. To maintain a controlled environment and facilitate observation of the participants, we will use the university's usability lab to conduct the study.

\subsection{Study Design}
The study will begin with an explanation of a labeling system, its development purpose, and the objectives of the experimental study. This information will be provided orally, allowing for questions and answers.
The experiment will be divided into two rounds: one for labeling rail faults by the train drivers and the other for labeling train vehicle faults by the workshop foremen. At the beginning of each round, participants will be given a text explanation outlining the goals and motivations behind the annotator user group. Following this, participants will be assigned tasks to complete on their own. As suggested by Nielsen \cite{ref_UsabilityEngineering}, we have designed the test tasks to be as representative as possible.
Tasks for the labeling of the train vehicles by the workshop foremen:
\begin{itemize}
    \item A train vehicle with the ID 918061439587DDB visited the workshop. You found a defect in the axle and repaired it. Now, you need to label the found and repaired defect in the system and submit the data to DigiOnTrack.
    \item Another train vehicle with the ID 918544650040CHBLS visited the workshop yesterday. You found two defects during maintenance, categorized as a flat spot and a commutator issue. Now, you must label the found and repaired defects in the labeling system and submit the data to DigiOnTrack.
\end{itemize}

Tasks for the labeling of the rails by the train drivers:
\begin{itemize}
    \item Two events occurred during today’s journey. The machine was very loud at around 1 pm near the Meyenburg train station, likely due to chatter marks. Later, at 4 pm, just before arriving at Putlitz station, you had to brake sharply because a deer was on the rails. Now, you must label these events and submit the data to DigiOnTrack.
    \item During your journey today, shortly after leaving Putlitz station, a loud bang indicated a rail breakage. You recall this event occurred around 8 pm. Now, you must label the event and submit the data to DigiOnTrack.
\end{itemize}

To eliminate any potential carryover effects, participants will be divided into two groups: one will start with the tasks designated for the workshop foreman, and the other will begin with the tasks assigned to the train drivers. The experiment is estimated to take 30 minutes. The next round will start either when participants have finished their tasks or when 10 minutes have passed. Participants will perform the round for labeling train vehicles by the workshop foremen on a desktop computer and the round for labeling rails by the train drivers on a tablet.

\subsection{Data Collection and Analysis} 
In addition to collecting demographic information such as age, gender, and occupation, we will use the Affinity for Technological Interaction scale \cite{ref_ATI} to assess participants’ affinity for technological interaction. After each round, usability will be measured using the System Usability Scale \cite{ref_SUSPaper}, and user experience will be measured using the User Experience Questionnaire \cite{ref_UEQ}. The System Usability Scale is a proven tool for measuring participants’ perceived usability \cite{ref_SUSPaper}, while the User Experience Questionnaire evaluates attractiveness, perspicuity, efficiency, dependability, stimulation, and novelty \cite{ref_UEQPaper}.
In addition to the questionnaires, participants’ performance will be measured by the time taken to complete tasks and the ratio of successful interactions to errors.
After completing the questionnaires in the second round, participants will undergo a debriefing session with the experimenter, during which they can comment on the system and suggest improvements.
Finally, the Spearmen-Correlation Test will be used to investigate whether there are correlations between age, affinity for technological interaction, and system usability or User Experience.

\section{Conclusion}
In this paper, we presented the design and implementation of a graphical labeling user interface integrated into a monitoring and predictive maintenance system for trains and rail infrastructure in a rural area of Germany. Our project aims to enhance rail transportation's economic viability and operational efficiency by utilizing cost-effective wireless monitoring systems, which combine affordable sensors and machine learning algorithms for predictive maintenance. By ensuring a high-quality labeling process through a user interface with high usability and user experience, we strive to integrate a successful and reliable predictive maintenance system.
The design of our labeling user interface was grounded in best practices of usability heuristics, specifically addressing the unique requirements in the field. By engaging closely with stakeholders, including railroad company employees, sensor system developers, and data exchange experts, we ensured that the labeling process was seamlessly integrated into the existing workflows of train drivers and workshop foremen. 

Currently, our project is at the stage where the labeling system has been technically implemented and is ready for usability and user experience testing. We have outlined a comprehensive study protocol to evaluate the system, aiming to validate our design approach and provide a useful reference for other researchers and developers working on labeling user interfaces for predictive maintenance systems in the rail transportation industry. Additionally, we aim to explore the scalability of our approach to other areas of infrastructure maintenance and extend the application of our system to broader contexts within the transportation industry.
The value of our work lies in its potential to significantly reduce maintenance costs and improve service reliability in rail transportation through a successfully integrated labeling system. By addressing the critical aspects of usability and user experience, we aim to ensure high-quality data labeling, which is essential for successfully deploying machine learning algorithms in predictive maintenance. Our approach not only contributes to the academic literature but also offers practical insights and solutions for the rail transportation sector, advancing the integration of Industry 4.0 technologies in this field.

\begin{credits}
\subsubsection{\ackname} We gratefully acknowledge financial support through the TÜV Rheinland Consulting GmbH with funds provided by the Federal Ministry for Digital and Transport (BMDV) under Grant No. 19F2265 (DigiOnTrack). We also want to thank the reviewers for their thoughtful feedback.

In this paper, we used Overleaf’s built-in spell checker, the current version of ChatGPT (GPT 3.5), and Grammarly. These tools helped us fix spelling mistakes and get suggestions to improve our writing. If not noted otherwise in a specific section, these tools were not used in other forms.

\subsubsection{\discintname}
The authors have no competing interests to declare relevant to this article's content. 
\end{credits}
%
%
%
 \bibliographystyle{splncs04}
 \bibliography{main}

\end{document}